\documentclass [preprint,showpacs,nofootinbib,superscriptaddress]{revtex4}
\usepackage{verbatim}
\usepackage{slashed}
\usepackage{epsfig}
\usepackage{feynmf}
\usepackage{subfigure}
\usepackage{graphicx}
\usepackage{amsmath}
\usepackage{mathrsfs}
\usepackage{bm}

\begin{document}

\preprint{DESY 13-009}

\title{The QED contribution to $J/\psi$ plus light hadrons production at B-factories}


\author{Zhi-Guo He}
\affiliation{
{\normalsize II. Institut f\"ur Theoretische Physik, Universit\"at Hamburg,}\\
{\normalsize Luruper Chaussee 149, 22761 Hamburg, Germany}
}
\author{Jian-Xiong Wang}

\affiliation{Institute of High Energy Physics, Chinese Academy of
Science, P.O. Box 918(4), Beijing, 100049, China.\\ Theoretical
Physics Center for Science Facilities,(CAS) Beijing, 100049, China.}

\date{\today}

\begin{abstract}\vspace{5mm}
To understand the direct $J/\psi+X_{\mathrm{non-c\bar{c}}}$ production mechanism in $e^{+}e^{-}$
annihilation, in this work, we propose to measure the inclusive $J/\psi$ plus light hadrons (LH)
production at B-factories and present a detailed study on its QED production due to
$\psi(2S)$ feed-down, where the $\psi(2S)$ are produced in $e^{+}e^{-}\to \psi(2S)+\gamma$ and
$e^{+}e^{-}\to \psi(2S)+f\bar{f},~f=lepton,light quark$, and QED contribution to direct $J/\psi+q\bar{q}$
production with $q=u,d,s$ quark. We find that the QED contribution is huge in the whole phase space region, but
can be reduced largely and is in the same order as the QCD contribution when
a suitable cut on the angel $\theta_{J/\psi}$ between $J/\psi$ and the $e^{+}e^{-}$ beam is made.
In this way, the cross section of $J/\psi+LH$ QCD production
can be obtained by subtracting the QED
contribution from the experimental measurement on inclusive $J/\psi$ plus light hadrons.
To help to remove the QED background,
we also calculate the angular and momentum distribution of $J/\psi$ in the QED contribution.

\end{abstract}


\maketitle
\section{Introduction}
The development of the non-relativistic QCD (NRQCD) effective field theory \cite{Bodwin:1994jh}
provides a powerful tool to study the production and decay of heavy quarkonium states that
are constituted by one heavy quark (Q) and one heavy anti-quark $\bar{Q}$. The virtual
difference between NRQCD and the conventional color-singlet model (CSM) is that it
allows the contribution of $Q\bar{Q}$ state in the color-octet (CO) configuration
at short-distance which finally evolves into heavy mesons through emission
of soft gluon(s) non-perturbatively. This is refereed as the CO mechanism (COM).
The role of the COM has been extensively studied in various high energy environments,
for reviews see Ref.\cite{Brambilla:2004wf}.

Among them, the $J/\psi$ production in $e^{+}e^{-}$ annihilation at B-factories (the Babar
and Belle) have attracted considerable solicitude in recent years. Experimentally, the
cross section for inclusive $J/\psi$ production was reported by the Babar \cite{Aubert:2001pd}
and Belle \cite{Abe:2001za} collaborations in 2001. Belle collaboration further divided
the inclusive $J/\psi$ production rate into two pieces: (a) $e^{+}e^{-}\to J/\psi+c\bar{c}$
part\footnote{Here the $J/\psi+c\bar{c}$ includes both the exclusive double charmonium
production process and the $J/\psi$ production process in association with charmed hadrons
, and the cross section for double charmonium production process was also reported by
Babar collaboration later \cite{Aubert:2005tj}.}, (b) $e^{+}e^{-}\to J/\psi+X_{\rm{non}
-c\bar{c}}$ part, and measured them separately \cite{Abe:2002rb,:2009nj}. The latest
results reported by the Belle are \cite{:2009nj}
\begin{subequations}
\begin{equation}
\sigma(e^{+}e^{-}\to J/\psi+X)=1.17\pm0.02\pm0.07\rm{pb},
\end{equation}
\begin{equation}
\sigma(e^{+}e^{-}\to J/\psi+c\bar{c}+X)=0.74\pm0.08^{+0.09}_{-0.08}\rm{pb},
\end{equation}
\begin{equation}
\sigma(e^{+}e^{-}\to J/\psi+X_{\rm{non-c\bar{c}}})=0.43\pm0.09\pm0.09\rm{pb}.
\end{equation}
\end{subequations}
In the case of $J/\psi+c\bar{c}$ production, where the the CO contribution is found to
be very small \cite{Liu:2003jj}, there were large discrepancies between Belle results
and NRQCD predictions at leading order (LO) in $\alpha_s$ and $v$
\cite{Cho:1996cg,Liu:2003jj,Braaten:2002fi}, where $v$ is the relative velocity between
$c$ and $\bar{c}$ in the meson rest frame. These puzzles are now largely resolved after
taking into account the next-to-leading order (NLO) QCD corrections
\cite{Zhang:2005cha,Gong:2007db} and the relativistic corrections \cite{Bodwin:2006ke}.
In contrast, in the $J/\psi+X_{\rm{non}-c\bar{c}}$ production case, the contribution of
the CO $e^{+}e^{-}\to c\bar{c}(^1S^{8}_0,\; ^3P^{8}_J)+g$
\cite{Braaten:1995ez,Yuan:1996ep,Fleming:2003gt,Wang:2003fw} process is expected to be
significant and even larger than that of the CS $e^{+}e^{-}\to c\bar{c}(^3S^{1}_1)+gg$
process. At LO, the cross sections of the CO and CS processes are predicted to be about
$0.3\sim0.8\rm{pb}$ and $0.2\sim0.3\rm{pb}$ \cite{Cho:1996cg,Driesen:1993us}, respectively.
Recently, the k-factor of their NLO QCD corrections were found to be about 1.3
\cite{Ma:2008gq,Gong:2009kp} and 1.9 \cite{Zhang:2009ym} correspondingly, and what's more
the relativistic corrections can also enhance the LO CS result by about $20\%-30\%$
\cite{He:2009uf,Jia:2009np}. Then up to the NLO of $\alpha_s$ and $v^2$, the cross
section of the CS contribution itself can reach about $440\sim560\rm{fb}$ \cite{He:2009uf}.
which almost saturate the Belle measurement and leave very little room for CO contribution.
So in $e^{+}e^{-}\to J/\psi+X_{\rm{non-c\bar{c}}}$ process, there is a large conflict
between NRQCD prediction and Belle current measurements. By setting the CS contribution
into zero, the upper limit of the CO matrix elements is obtained \cite{Zhang:2009ym}
\begin{equation}\label{CO1}
\langle0|\mathcal{O}(^1S_0^{8})|0\rangle^{J/\psi}+4.0 \langle0|\mathcal{O}(^3P_J^{8})
|0\rangle^{J/\psi}/m_c^2<(2.0\pm0.6)\times10^{-2}\rm{GeV}^{3}.
\end{equation}

However, in some other processes of $J/\psi$ production, the recent theoretical
calculations shown that the CO contribution is important. For example, (a) for
$J/\psi$ production from $Z$ decay, the CS result at QCD NLO \cite{Li:2010xu}, can
only account for one-half of the experimental data and the other half might be
attributed to the CO contribution; (b) for $J/\psi$ production in $\Upsilon$ decay,
there is large gap between CS contribution \cite{He:2009by} and the experimental result;
(c) for $J/\psi$ photoproduction at HERA, the transverse momentum ($p_t$) distribution
and the polarization parameters of $J/\psi$ can not be well described by the CS channel
at QCD NLO alone as well \cite{Artoisenet:2009xh,Chang:2009uj}, and the NRQCD prediction
that includes both the CO and CS contributions can give a well description of the $J/\psi$
$p_t$ distribution when the NLO QCD corrections are taken into account
\cite{Butenschoen:2009zy}; (d) for $J/\psi$ hadroproduction, despite of the huge NLO QCD
corrections \cite{Campbell:2007ws,Gong:2008sn,Gong:2008hk}, the CS contribution still
can not explain the experimental measurements, and the role of the COM is significant
\cite{Braaten:1994vv,Gong:2008ft,Ma:2010yw,Butenschoen:2010rq}. By fitting the $J/\psi$
hadroproduction data with the complete NRQCD results at QCD NLO, including both the CS
and CO contributions, two different sets of constraint for the CO matrix elements are
obtained, which are $\langle0|\mathcal{O}(^1S_0^{8})|0\rangle^{J/\psi}+ 3.9
\langle0|\mathcal{O}(^3P_J^{8})|0\rangle^{J/\psi}/m_c^2=7.4\times10^{-2}\rm{GeV}^{3}$
\cite{Ma:2010yw}, and $\langle0|\mathcal{O}(^1S_0^{8})|0\rangle^{J/\psi}+ 3.9
\langle0|\mathcal{O} (^3P_J^{8})|0\rangle^{J/\psi}/m_c^2=2.4\times10^{-2}\rm{GeV}^{3}$
\cite{Butenschoen:2010rq}. Although these two results are not consistent with
each other, both of them exceed the upper limit given in Eq.(\ref{CO1}). In particular,
the former one is three times larger than the limit in Eq.(\ref{CO1}). These studies
yield almost completely opposite conclusion about how large the CO contribution is, or in
other words, how large the values of the CO matrix elements could be.

After comparing the results of the Babar and Belle with the theoretical calculation meticulously,
we find that there are some uncertainties which can potentially have large impact on the
current conclusion. One is that the Babar measurement on $J/\psi$ inclusive production cross section
\cite{Aubert:2001pd} is about two times larger than that of Belle\cite{:2009nj}. If we subtract
the measurement of the Belle $\sigma(J/\psi+c\bar{c}+X)=0.74\rm{pb}$, which is well understood
theoretically, from Babar result, there will be enough room left for CO contribution. One
possible reason for the different results of the Babar and Belle is that they use different
methods to select the data. Another uncertainty is that,in the latest measurement of the Belle
\cite{:2009nj}, they only select the event that includes at least five charge tracks in the final
states, and make no corrections. This means that all events that include zero or two charged
light hadrons, such as $J/\psi+m(\pi^{+}\pi^{-})+n\pi^{0}$ for $(m=0,1;n=0,1,2\ldots)$, are
excluded. From the point view of quark-hadron duality, Belles measurements do not include the
whole NRQCD predictions. It may has little influence on the measurement of $\sigma (e^{+}e^{-}
\to J/\psi+c\bar{c})$ \cite{:2009nj}, but large influence on that of $\sigma(e^{+}e^{-}\to
J/\psi+X_ {\rm{non}-c\bar{c}})$ from the non-perturbative hadronization mechanism of gluons.
To reduce the uncertainties mentioned above and understand the $J/\psi+X_ {\rm{non}-c\bar{c}}$
production mechanism, we suggest to measure the cross section of $J/\psi+\mathrm {light
\;hadrons\;(LH)}$ production by the Belle and Babar collaborations with the same kinematic criteria,
which can be compared with the theoretical prediction directly.

Besides the interesting conventional QCD contribution, there are also large QED backgrounds
due to $\psi(2S) \to J/\psi+\pi\pi$\footnote{$\psi(2S)$ can also decay into $J/\psi+\eta$.
However the branching ratio is more 15 times smaller than the $2 \pi$ channel, so we do
not take it into account in our calculation.}, where $\psi(2S)$ is produced in the initial
state radiation (ISR) process $e^{+}e^{-}\to \psi(2S)+\gamma$ and higher order QED processes
$e^{+}e^{-}\to \psi(2S)+f\bar{f}$ ($f$ can be lepton or light quark), and direct $J/\psi$
production in the $e^{+}e^{-}\to 2\gamma^{\ast}\to J/\psi+q\bar{q}$ process with $q=u,d,s$
quark. To help to remove them, in this work, we will present a detailed study about the $\psi(2S)$
and $J/\psi+q\bar{q}$ productions in the QED processes and their influence on the $J/\psi+\rm{LH}$
measurement.

\section{Framework of Calculation}

  For the process of $J/\psi+\pi\pi$ production from $\psi(2S)$ feed-down, the Feynman
amplitude $\mathcal{M}$ can be generally written as:
\begin{equation}
\mathcal{M}=\mathcal{M}_{\mu}^{\psi(2S)}(P_{2S})\times\frac{-g^{\mu\nu}+\frac{P_{2S}^{\mu}P_{2S}^{\nu}}
{P_{2S}^2}}{P_{2S}^2-M_{2S}^2+i*M_{2S}\Gamma}\mathcal{M}_{\nu}^{(\psi(2S)\to J/\psi+\pi\pi)}
\end{equation}
where $\mathcal{M}_{\mu}^{\psi(2S)}(P_{2S})$ and $\mathcal{M}_{\nu}^{(\psi(2S)\to J/\psi+\pi\pi)}$
are the Feynman amplitudes for $\psi(2S)$ production with momentum $P_{2S}$ and $\psi(2S)$
decay into $J/\psi+\pi\pi$ respectively, and $\Gamma$ is the total decay width of $\psi(2S)$.
Using narrow width approximation
\begin{equation}
\lim_{\Gamma \to 0} \frac{1}{(P_{2S}^2-M_{2S}^2)^2+M_{2S}^2\Gamma^2}=\frac{\pi\delta(P_{2S}^2-M_{2S}^2)}{M_{2S}\Gamma},
\end{equation}
it is straightforward to obtain the expression for the corresponding cross section which is
factorized as the product of the cross section of $\psi(2S)$ production and the branching
function of $\psi(2S)\to J/\psi+\pi\pi$:
\begin{equation}\label{cross}
\sigma=\frac{1}{8s}\int\sum|\mathcal{M}^{\psi(2S)}|^2\;d\mathrm{LIPS_1} \times
\frac{1}{2(2J+1)M_{2S}\Gamma}\int\sum|\mathcal{M}^{(\psi(2S)\to J/\psi+\pi\pi)}|^2\; d\mathrm{LIPS}_2,
\end{equation}
where $\mathrm{LIPS_1}$ is the phase space of $\psi(2S)$ production, $\mathrm{LIPS}_2$ is
the phase space of $\psi(2S)$ decay into $J/\psi+\pi\pi$, and $J=1$ is the spin of $\psi(2S)$.

We use the effective Lagrangian that is constructed in Ref.\cite{Mannel:1995jt} to describe
$\psi(2S)\to J/\psi+\pi\pi$. The amplitude $\mathcal{M}^{(\psi(2S)\to J/\psi+\pi\pi)}$
can be read directly from the Lagrangian
\begin{eqnarray}\label{para_0}
\mathcal{M}^{(\psi(2S)\to J/\psi+\pi(p_1)\pi(p_2))}=-\frac{4}{F_0^2}\Big{[}\left( \frac{g}{2}
(m_{\pi\pi}^2-2M_{\pi}^2)+g_1(v\cdot p_{1})(v\cdot p_{2})+g_3 M_{\pi}^2\right)\nonumber\\
\times
\epsilon^{\ast}_{J/\psi}\cdot \epsilon_{\psi(2S)}
 +g_2(p_{1\mu}p_{2\nu}+p_{1\nu}p_{2\mu})
\epsilon^{\ast\mu}_{J/\psi}\epsilon^{\nu}_{\psi(2S)}\Big{]}
\end{eqnarray}
where $m^{2}_{\pi\pi}=(p_1+p_2)^2$, $M_{\pi}$ is the mass of $\pi$ meson, and  $v=(1,\vec{0})$
in the rest frame of $\psi(2S)$. In their convention, the $\pi$ decay constant $F_0\simeq93MeV$.
The coupling constant $g_2\simeq0$, because it is strongly suppressed by the chiral symmetry
breaking scale over $m_c$. By fitting the distributions of $m_{\pi\pi}$ and $\cos\theta_{\pi}^{\ast}$,
which is the angel between $J/\psi$ and $\pi^{+}$ in the rest frame of $\psi(2S)$, in the
decay of $\psi(2S)\to J/\psi+\pi^{+}\pi^{-}$, the BES Collaboration obtained two set results for
$\frac{g_1}{g}$ and $\frac{g_3}{g}$ \cite{Bai:1999mj}. Together with
$\mathrm{Br}(\psi(2S)\to J/\psi+\pi^{+}\pi^{-})=33.6\%$ \cite{Nakamura:2010zzi}, they then
obtained that\footnote{These parameters can also well reproduce the decay width of
$\psi(2S)\to J/\psi+\pi^{0}\pi^{0}$.}
\begin{equation}\label{para_1}
g=0.322,\;\frac{g_1}{g}=-0.49,\;\frac{g_3}{g}=0.54,
\end{equation}
or
\begin{equation}\label{para_2}
g=0.319,\;\frac{g_1}{g}=-0.347,\; g_3=0.
\end{equation}

For the processes considered, $\mathcal{M}^{(\psi(2S)\to J/\psi+\pi\pi)}$ is common,
so we essentially only need to compute $\mathcal{M}^{2S}$. In the non-relativistic limit, for
the QED process of $e^{+} e^{-}\to \psi(2S)+X$ the factorization formula in CSM and NRQCD are
equivalent, and the amplitude $\mathcal{M}^{\psi(2S)}$ can be written as:
\begin{equation}
\mathcal{M}^{\psi(2S)}=\sqrt{C_{2S}}\sum_{s_1,s_2}\sum_{i,j}\langle s_1;s_2|1\;S_z\rangle
\langle 3i;\bar{3}j|1\rangle \mathcal{M}(e^{+}e^{-}\to c_i(\frac{P_{2S}}{2},s_1)+\bar{c}_{j}
(\frac{P_{2S}}{2},s_2)+X)
\end{equation}
where $\mathcal{M}$ is the standard Feynman amplitude for $e^{+}e^{-}\to c_{i}(\frac{P_{2S}}{2},s_1)+\bar{c}_{j}(\frac{P_{2S}}{2},s_2)+X$,
$\langle 3i;\bar{3}j|1\rangle=1/\sqrt{N_c}$ and $\langle s_1;s_2|1\;S_z\rangle$
are the SU(3)-color and SU(2)-spin Clebsch-Gordan coefficients for $c\bar{c}$ projecting on
the CS spin-triplet S-wave state. The projection of Dirac spinors can be re-expressed as:
\begin{equation}
\sum_{s_1,s_2} \langle s_1;s_2|1\;S_z\rangle v(\frac{P_{2S}}{2},s_2)\bar{u}(\frac{P_{2S}}{2},s_1)
=\frac{1}{2\sqrt{2}}\slashed{\epsilon}^{\ast}(S_z)(P_{2S}+M_{2S}).
\end{equation}
$C_{2S}$ can be related to the $\psi(2S)$ wave function at origin by $C_{2S}=\frac{1}{4\pi}|R_{2S}(0)|$.
And $|R_{2S}(0)|$ can be obtained from potential model calculation or can be determined from $\psi(2S)$
decay into $e^{+}e^{-}$ with
\begin{equation}
\Gamma(\psi(2S)\to e^{+}e^{-})=\frac{16\alpha^2|R_{2S}(0)|^2}{9M_{2S}^2}
\end{equation}

\section{The Feed-down Background From $e^{+}e^{-}\to \psi(2S)+\gamma$ }

\begin{figure}
\begin{center}
\includegraphics[scale=0.8]{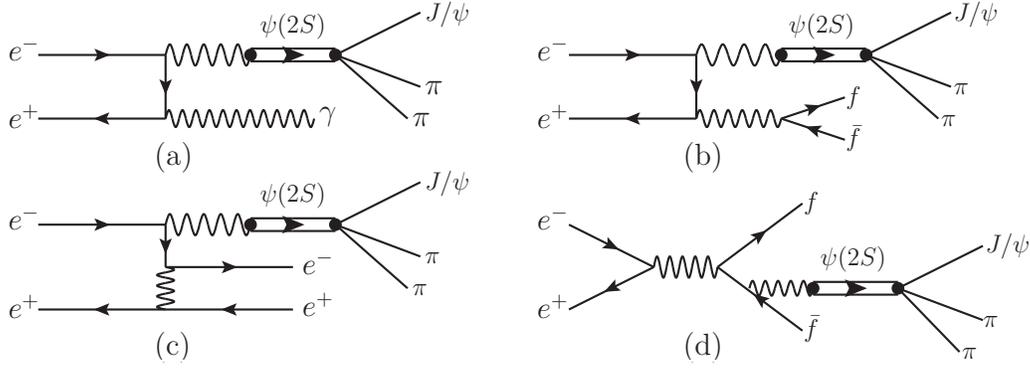}
\caption{ The typical diagrams describing the $\psi(2S)$ feed-down background. }
\end{center}
\end{figure}

  The typical Feynman diagrams for the ISR process $e^{+}e^{-}\to \psi(2S)+\gamma$ followed
by $\psi(2S)\to J/\psi+\pi\pi$ are shown in Fig.(1a). Using the formula introduced in Eq.(5-8),
we compute $|\mathcal{M}(e^{+}e^{-}\to \psi(2S)+\gamma)|^2$ analytically and obtain
\begin{eqnarray}\label{ISR}
&&|\mathcal{M}(e^{+}e^{-}\to \psi(2S)+\gamma)|^2=
\frac{96e_c^2(4\pi\alpha)^3C_{2S}}{s\, r\,{\left( -1 + r +
        \left( -1 + r \right) \,\left( -1 + 4\,r_e \right) \,
         x_2^2 \right) }^2}
\Big{(} -1 - 2\,r - r^2 \nonumber\\
&&- 8\,r_e +
      16\,r\,r_e + 32\,r_e^2 -
      4\,\left( r + 2\,r_e \right) \,
       \left( -1 + 4\,r_e \right) \,x_2^2 +
      {\left( -1 + r \right) }^2\,
       {\left( 1 - 4\,r_e \right) }^2\,x_2^4
      \Big{)}
\end{eqnarray}

where $e_c=\frac{2}{3}$, $r=\frac{M_{2S}^2}{s}$,  $r_e=\frac{M_{e}^2}{s}$,
$x_2=\cos(\theta_{\psi(2S)})$ and $\theta_{\psi(2S)}$ is the angel between $\psi(2S)$
and the $e^{+}e^{-}$ beam. In the
limit of $r_e=0$, Eq.(\ref{ISR}) can be simplified as:
\begin{equation}\label{Coll}
|\mathcal{M}(e^{+}e^{-}\to \psi(2S)+\gamma)|^2=\frac{96e_c^2(4\pi\alpha)^3C_{2S}}{s}
\big{(}\frac{1}{r}-\frac{2(1+r^2)} {r(1-r)^2(1-x_{2}^{2})}\big{)}.
\end{equation}

Setting $M_{2S}=3.686\mathrm{GeV}$, $m_e=0.51\mathrm{MeV}$, $\alpha=\frac{1}{137}$,
and using $\Gamma(\psi(2S)\to e^{+}e^{-})=4.30\mathrm{keV}$, we get
\begin{equation}
\sigma(e^{+}e^{-}\to \psi(2S)+\gamma)=13.22\mathrm{pb}.
\end{equation}
And the feed-down production
\begin{equation}\label{ISR2}
\sigma(e^{+}e^{-}\to \psi(2S)+\gamma)\times Br(\psi(2S)\to J/\psi+\pi\pi)=6.79\mathrm{pb},
\end{equation}
which, as expected, is huge. This is because in the limit of $m_{e}\rightarrow 0$,
there will be collinear singularities in $|\mathcal{M}(e^{+}e^{-}\to \psi(2S)+\gamma)|^2$
in Eq.(\ref{Coll}) at $x_2=\pm1$ points. The angular distribution
$\frac{d\sigma(e^{+}e^{-}\to \psi(2S)+\gamma)}{dx_2}$ is shown in Fig.[2]. It can be found from
Fig.[2] that the differential cross section drops down very fast when $\psi(2S)$ goes off the beam line
a little. If we make a cut on $x_2$, i.e the angle $\theta_{\psi(2S)}$, the cross
section will be reduced largely. The cross sections in different cut conditions are given below:
\begin{subequations}\label{result1}
\begin{equation}
\sigma(e^{+}e^{-}\to \psi(2S)+\gamma)\times Br(\psi(2S)\to J/\psi+\pi\pi)\Big{|}_{\frac{\pi}{18}<\theta_{\psi(2S)}<\frac{17\pi}{18}}=
1.48\mathrm{pb},
\end{equation}
\begin{equation}
\sigma(e^{+}e^{-}\to \psi(2S)+\gamma)\times Br(\psi(2S)\to J/\psi+\pi\pi)\Big{|}_{\frac{\pi}{9}<\theta_{\psi(2S)}<\frac{8\pi}{9}}=
0.99\mathrm{pb},
\end{equation}
\begin{equation}
\sigma(e^{+}e^{-}\to \psi(2S)+\gamma)\times Br(\psi(2S)\to J/\psi+\pi\pi)\Big{|}_{\frac{\pi}{6}<\theta_{\psi(2S)}<\frac{5\pi}{6}}=
0.71\mathrm{pb}.
\end{equation}
\end{subequations}

\begin{figure}
\begin{center}
\includegraphics[scale=0.8]{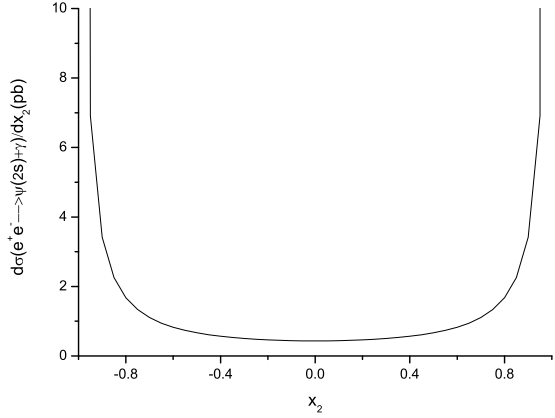}
\caption{ The angular distribution of $\psi(2S)$ in the ISR process $e^{+}e^{-}\to \psi(2S)+\gamma$,
where $x_2=\cos(\theta_{\psi(2S)})$, and $\theta_{\psi(2S)}$ is the angular between $\psi(2S)$
and the $e^{+}e^{-}$ beam.}
\end{center}
\end{figure}

Let $p^{\ast\mu}_{J/\psi}$ denotes the four-momentum of $J/\psi$ in the rest frame of
$\psi(2S)$,  then $|\vec{\mathbf{p}}^{\ast}_{J/\psi}|/E^{\ast}_{J/\psi}$, the three-velocity
of $J/\psi$, ranges from 0 to 0.15, which is much smaller than that  of $\psi(2S)$
in the center of mass frame (CMF) of $e^{+}e^{-}$ collision, which is about 0.78. So the
angular distribution of $J/\psi$ can be obtained approximately by setting $\theta_{\psi(2S)}=\theta_{J/\psi}$,
where $\theta_{J/\psi}$ is the angel between $J/\psi$ and the $e^{+}e^{-}$ beam.
However, such an approximation may not be good enough here, because the cross section of $J/\psi+\pi\pi$
produced from the feed-down of $\psi(2S)$ ISR process (Eq.(\ref{ISR2})) is more than 10 times
larger than that of the CS QCD process\cite{He:2009uf}, and a tiny difference may potentially
result in a considerable effect. In this work, we calculated it directly. In the CMF,
$x_2^{\prime}=\cos(\theta_{J/\psi})$, can be expressed as:
\begin{equation}
x_{2}^{\prime}=\frac{\mathbf{k}_{1} \cdot {{\mathbf{p}}^{\prime}_{J/\psi}}}
{|\mathbf{k}_1||\mathbf{p}^{\prime}_{J/\psi}|},~
p^{\prime}_{J/\psi}=L p^{\ast}_{J/\psi}
\end{equation}
where $k_{1}$ is the four-momentum of $e^{+}$, $L$ is the Lorentz transformation from $\psi(2s)$ rest frame to
the CMF.  $p^{\prime}_{J/\psi}$ is the $J/\psi$ four-momentum in the CMF.
To do the calculation, the formula for the decay $\psi(2s)\rightarrow J/\psi + \pi\pi$ and
$\psi(2s)$ production are placed in the numerical phase space integration program generated by using
the Feynman Diagram Calculation (FDC) package \cite{Wang:2004du}, in which the Lorentz transformation
and the cut conditions  are employed in the numerical calculation.
The decay $\psi(2s)\rightarrow J/\psi + \pi\pi$ is calculated by using Eq.(\ref{para_0})
with $M_{\pi^{+}}=M_{\pi^{-}}=140\mathrm{MeV}$ and $M_{\pi^{0}}=135\mathrm{MeV}$.
When parameter set in Eq.(\ref{para_1}) are used,
the cross sections in different cut conditions are
\begin{subequations}\label{result2}
\begin{equation}
\sigma(e^{+}e^{-}\to (J/\psi+\pi\pi)_{\psi(2S)}+\gamma)
\Big{|}_{\frac{\pi}{18}<\theta_{J/\psi}<\frac{17\pi}{18}}=1.51\mathrm{pb};
\end{equation}
\begin{equation}
\sigma(e^{+}e^{-}\to (J/\psi+\pi\pi)_{\psi(2S)}+\gamma)
\Big{|}_{\frac{\pi}{9}<\theta_{J/\psi}<\frac{8\pi}{9}}=0.99\mathrm{pb};
\end{equation}
\begin{equation}
\sigma(e^{+}e^{-}\to (J/\psi+\pi\pi)_{\psi(2S)}+\gamma)
\Big{|}_{\frac{\pi}{6}<\theta_{J/\psi}<\frac{5\pi}{6}}=0.71\mathrm{pb}.
\end{equation}
\end{subequations}
Alternatively, if we choose the parameter set in Eq.(\ref{para_2}), the corresponding
cross sections become:
\begin{subequations}\label{result3}
\begin{equation}
\sigma(e^{+}e^{-}\to (J/\psi+\pi\pi)_{\psi(2S)}+\gamma)
\Big{|}_{\frac{\pi}{18}<\theta_{J/\psi}<\frac{17\pi}{18}}=1.52\mathrm{pb};
\end{equation}
\begin{equation}
\sigma(e^{+}e^{-}\to (J/\psi+\pi\pi)_{\psi(2S)}+\gamma)
\Big{|}_{\frac{\pi}{9}<\theta_{J/\psi}<\frac{8\pi}{9}}=1.00\mathrm{pb};
\end{equation}
\begin{equation}
\sigma(e^{+}e^{-}\to (J/\psi+\pi\pi)_{\psi(2S)}+\gamma)
\Big{|}_{\frac{\pi}{6}<\theta_{J/\psi}<\frac{5\pi}{6}}=0.71\mathrm{pb}.
\end{equation}
\end{subequations}
The numerical results in Eq.(\ref{result1},\ref{result2},\ref{result3}) show
that for $J/\psi+\pi\pi$ production from the ISR $\psi(2S)$ feed-down process
the approximation
\begin{equation}
\frac{d\sigma(e^{+}e^{-}\to (J/\psi+\pi\pi)_{\psi(2S)}+\gamma)}{d\cos(\theta_{J/\psi})}
=\frac{d\sigma(e^{+}e^{-}\to \psi(2S)+\gamma)\times Br(\psi(2S)\to J/\psi+\pi\pi)}
{d\cos(\theta_{\psi(2S)})}
\end{equation}
holds very well in the range of $\pi/9<\theta_{J/\psi}<8\pi/9$ at $\sqrt{s}=10.6$ GeV, and
the $J/\psi$ angular distribution is almost not dependent on the details about
how $\psi(2S)$ decays into $J/\psi+\pi\pi$. Hence the angular distribution of $J/\psi$ can
be safely obtained by using the angular distribution of $\psi(2S)$ in the interval
$\pi/9<\theta_{J/\psi}<8\pi/9$ with an additional renormalization factor of branching ratio of $\psi(2S)\to J/\psi+\pi\pi$.

\begin{figure}
\begin{center}
\subfigure[]
{\includegraphics[width=0.48\textwidth]{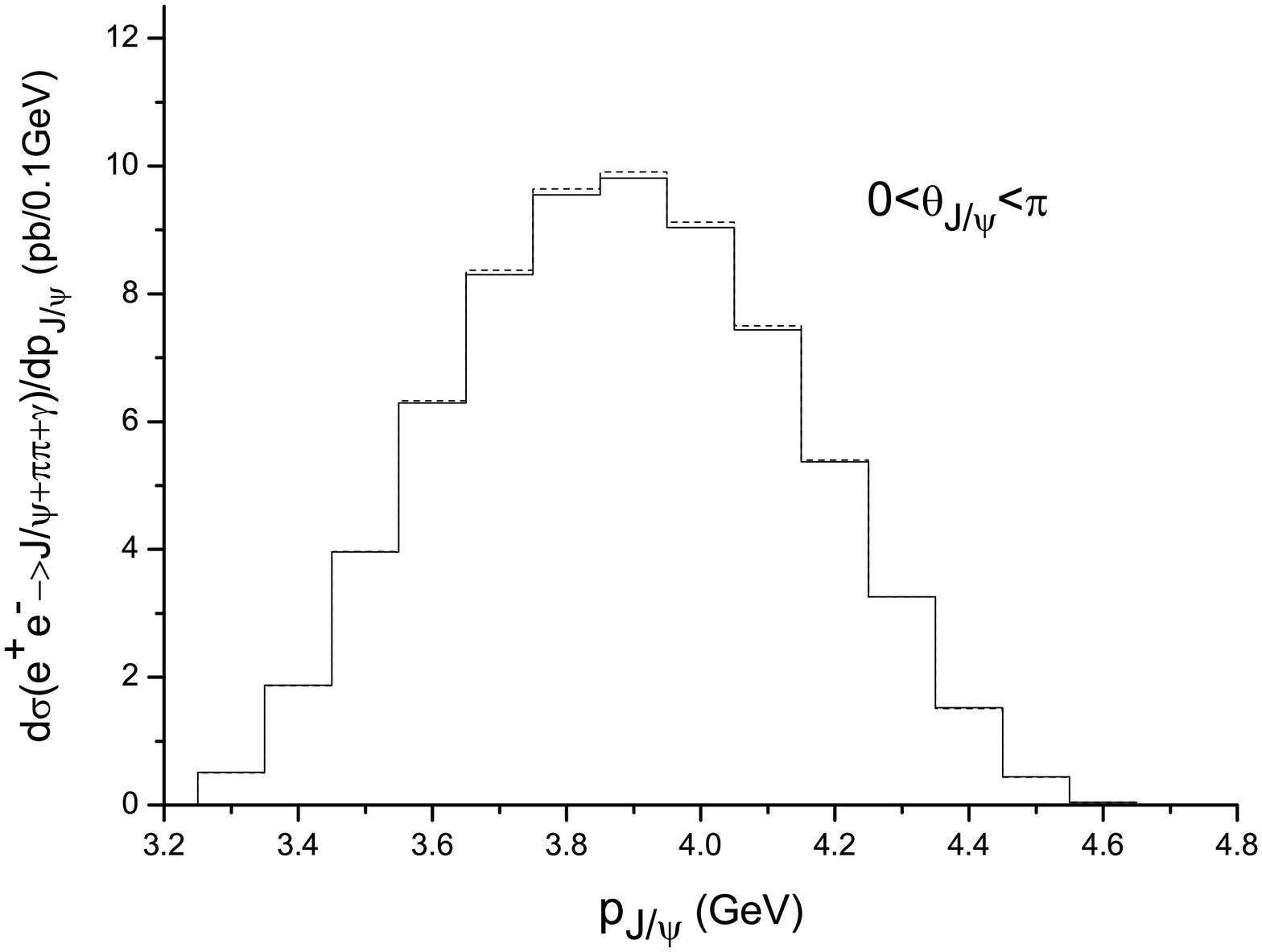}\label{fig:res_a}}
\subfigure[]
{\includegraphics[width=0.48\textwidth]{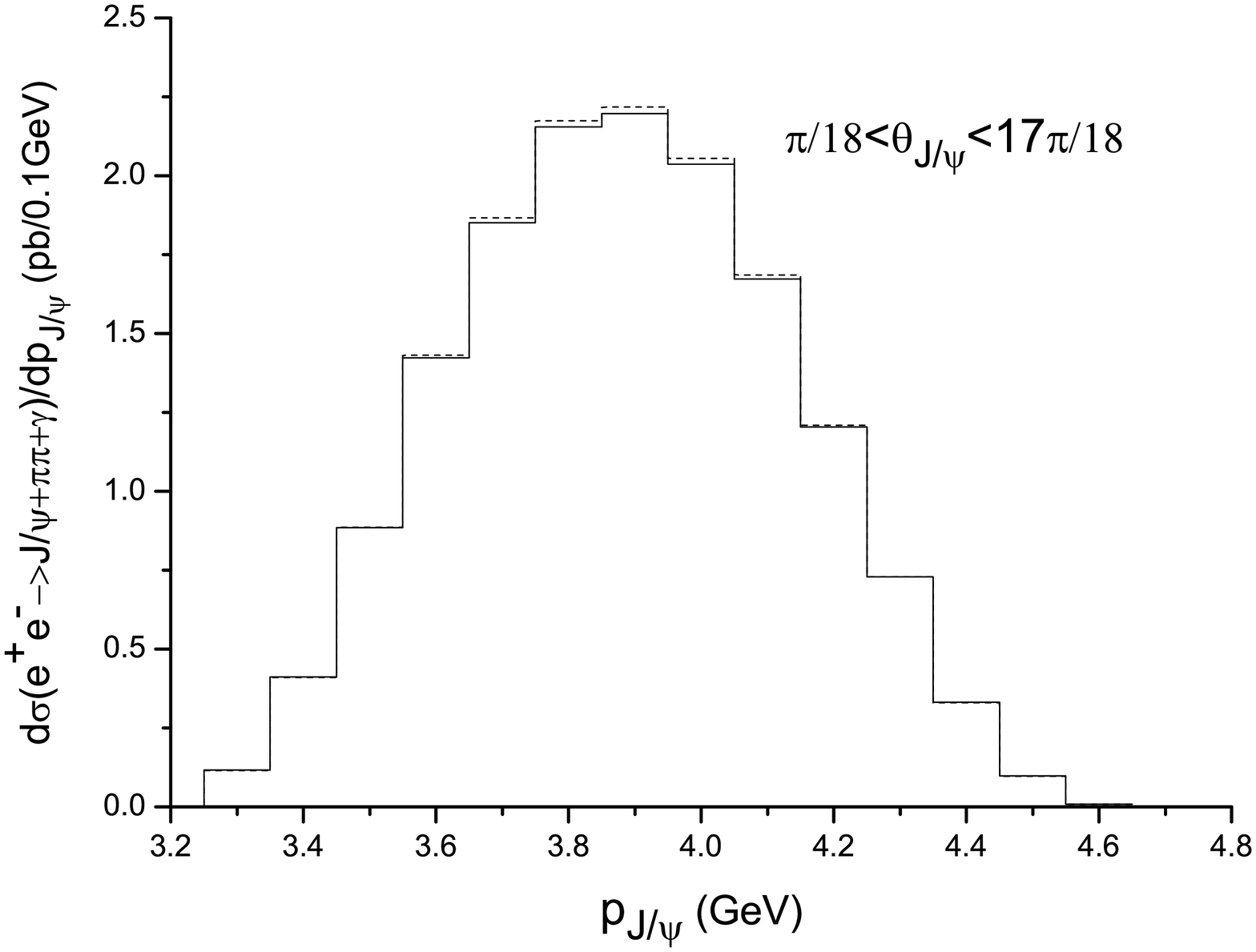}\label{fig:res_b}}
\subfigure[]
{\includegraphics[width=0.48\textwidth]{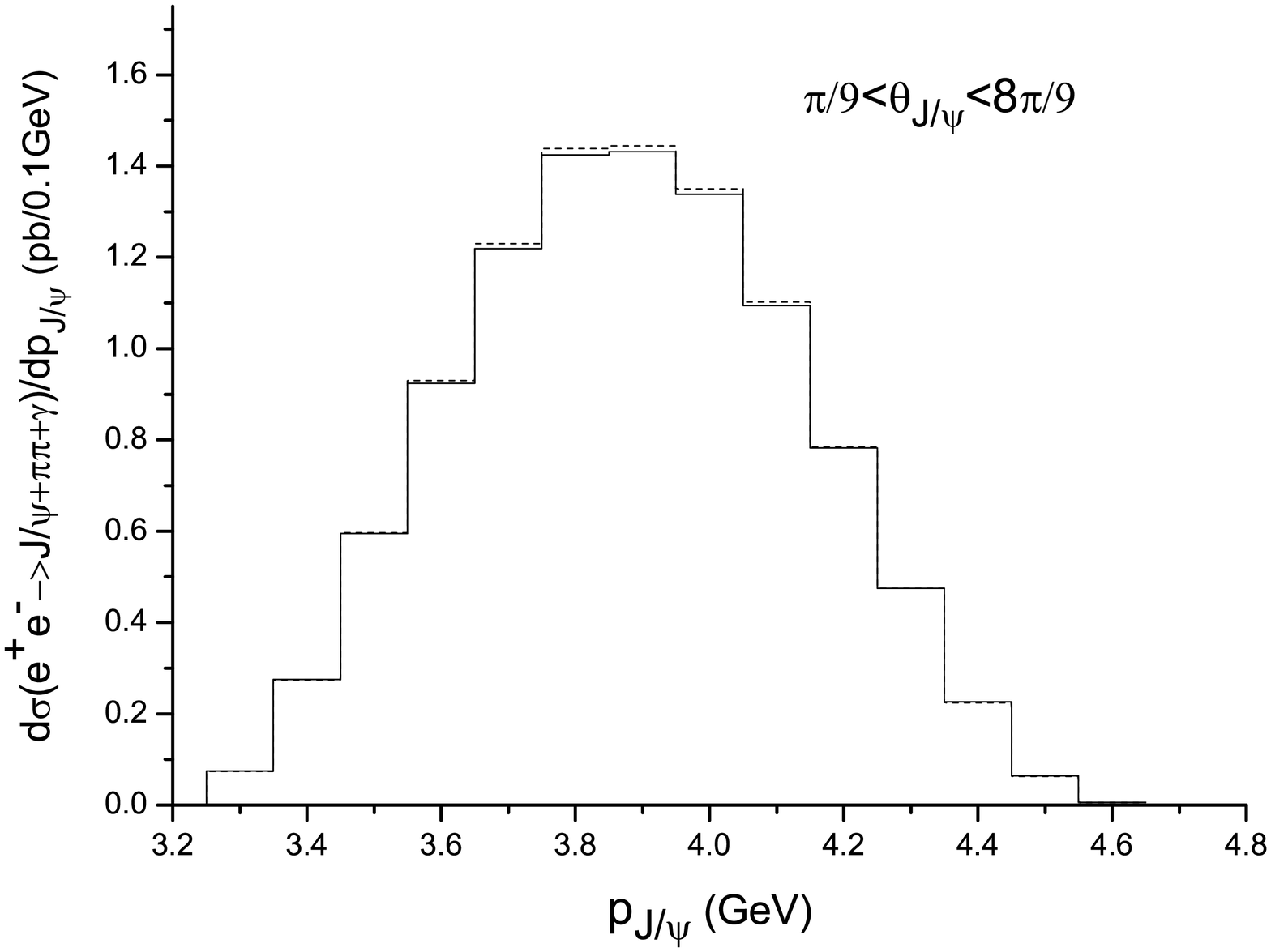}\label{fig:res_c}}
\subfigure[]
{\includegraphics[width=0.48\textwidth]{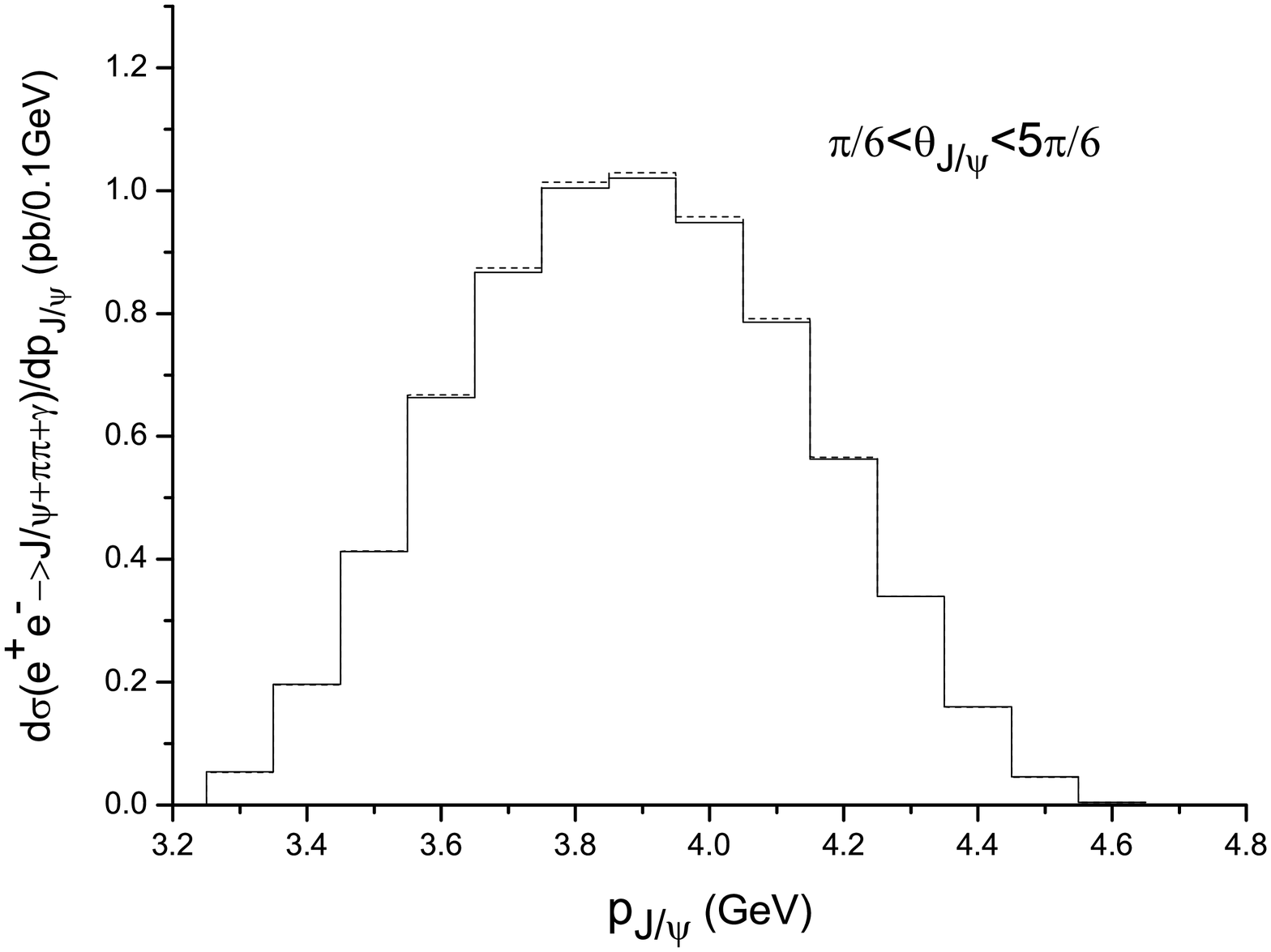}\label{fig:res_d}}
\caption{ The momentum spectra of $J/\psi$ produced from the feed-down of ISR $\psi(2S)$
process in different cut condition of $\theta_{J/\psi}$ by using two different
sets of parameters in Eq.(\ref{para_1}) (solid line) and Eq.(\ref{para_2}) (dashed line)
to describe $\psi(2S)\to J/\psi+\pi\pi$.}
\end{center}
\end{figure}

Because the energy difference between $\psi(2S)$ and $J/\psi$ is at the same order as
the energy of the soft gluon emitted from the CO $c\bar{c}(^{3}P_{J}^{8},^{1}S_{0}^{8})$
states \cite{Bodwin:1994jh}, which is of $m_cv^2$ order, there is a large overlap between
the kinematic region of the $J/\psi$ coming from ISR $\psi(2S)$ feed-down and that of the
$J/\psi$ produced in the CO process. To measure the CO $J/\psi$ production, it is helpful
to know the momentum distribution of $J/\psi$ production in the feed-down from the ISR $\psi(2S)$ process. We calculate
it numerically with different cut conditions of $\theta_{J/\psi}$ by using the two set of
parameters in Eq.(\ref{para_1},\ref{para_2}), and the results are given in Fig.[\ref{fig:res_a}-\ref{fig:res_d}].
The results in Fig.[\ref{fig:res_a}-\ref{fig:res_d}] show that similar to the angular
distribution, the momentum spectra give almost same results for those two different parameter sets.
Hence for simplicity, we only choose the parameter set
in Eq.(\ref{para_1}) in the following calculations.

\section{ Background From Higher QED Processes }

In direct $J/\psi$ production, the background coming from higher QED processes $e^{+}e^{-} \to
J/\psi+f+\bar{f}$ is also considerable \cite{Chang:1998pz}, where $f$ can be lepton or
light quark, and therefore the feed-down background from $e^{+}e^{-} \to \psi(2S)+f+\bar{f}$
can not be ignored too. The typical Feynman diagrams for $f\neq e$ are shown in Fig.(1b)
and Fig.(1d). When $f=e$, there are additional t-channel diagrams, the typical one of which
is shown in Fig.(1c). Because of this t-channel enhancement, the cross section for $f=e$ is
expected to be much larger than $f\neq e$ case. We will discuss $f=e$ and $f\neq e$ cases
separately in the subsections. At this order, in addition to the $\psi(2S)$ feed-down, there
is also sizable QED contribution from direct $J/\psi+q\bar{q}$ production with $q=u,d,s$ quark,
about which we will discuss in subsection C.

\subsection{The Feed-Down Background From $e^{+}e^{-} \to \psi(2S)+e^{+}e^{-}$ }

According to the interaction type of the initial $e^{+}e^{-}$, we divide the $e^{+}e^{-} \to
\psi(2S)+e^{+}e^{-}$ process into three part: the t-channel part (Fig.(1c)), the two-photon
channel part (Fig.(1b)), and the s-channel part (Fig.(1d)). It is easy to check that the
Feynman amplitude for each part itself is gauge invariant. Compared to cross section
$\sigma^{\mathrm{T}}$ for the t-channel part, the cross sections for the two-photon part
$\sigma^{\mathrm{TP}}$ and the s-channel part $\sigma^{\mathrm{S}}$ are suppressed by the
factors $\frac{M_{\psi(2S)}^2}{s}$, and $\frac{M_{\psi(2S)}^2}{s}\ln^{-2}(\frac{s}{4M_e^2})$
respectively, which are about $10^{-1}$ and $10^{-4}$ orders accordingly at $\sqrt{s}=10.6 \mathrm{GeV}$.
Choosing the same values for the parameters as in the ISR process, we obtained
\begin{subequations}\label{result4}
\begin{equation}
\sigma^{\mathrm{T}}(e^{+}e^{-}\to (J/\psi+\pi\pi)_{\psi(2S)}+e^{+}e^{-})=0.50\;\mathrm{pb};
\end{equation}
\begin{equation}
\sigma^{\mathrm{TP}}(e^{+}e^{-}\to (J/\psi+\pi\pi)_{\psi(2S)}+e^{+}e^{-})=4.8\times10^{-2}\;\mathrm{pb};
\end{equation}
\begin{equation}
\sigma^{\mathrm{S}}(e^{+}e^{-}\to (J/\psi+\pi\pi)_{\psi(2S)}+e^{+}e^{-})=8.5\times10^{-4}\;\mathrm{pb}.
\end{equation}
\end{subequations}
which are consistent with the qualitative estimation. The contribution of the s-channel part
is only $\sim1\;\mathrm{fb}$ order, which is about three times order less that the t-channel
contribution, so we drop it in the later analysis.

If we make the same cut on the $\theta_{J/\psi}$, $\sigma^{\mathrm{T}}$ and $\sigma^{\mathrm{TP}}$
both drop down largely too:
\begin{subequations}\label{result5}
\begin{equation}
\sigma^{\mathrm{T(TP)}}(e^{+}e^{-}\to (J/\psi+\pi\pi)_{\psi(2S)}+e^{+}e^{-})
\Big{|}_{\frac{\pi}{18}<\theta_{J/\psi}<\frac{17\pi}{18}}=0.11(0.019)\mathrm{pb};
\end{equation}
\begin{equation}
\sigma^{\mathrm{T(TP)}}(e^{+}e^{-}\to (J/\psi+\pi\pi)_{\psi(2S)}+e^{+}e^{-})
\Big{|}_{\frac{\pi}{9}<\theta_{J/\psi}<\frac{8\pi}{9}}=0.059(0.013)\mathrm{pb};
\end{equation}
\begin{equation}
\sigma^{\mathrm{T(TP)}}(e^{+}e^{-}\to (J/\psi+\pi\pi)_{\psi(2S)}+e^{+}e^{-})
\Big{|}_{\frac{\pi}{6}<\theta_{J/\psi}<\frac{5\pi}{6}}=0.039(0.010)\mathrm{pb}.
\end{equation}
\end{subequations}
The angular distribution of $\psi(2S)$ in the t- and two-photon channel parts are shown
in Fig.[4]. We find that in $\pi/9<\theta_{J/\psi}<8\pi/9$ region the
angular distribution of the $J/\psi$ production from feed-down can be obtained by using
that of $\psi(2S)$ as well with an additional renormalization factor of
$Br(\psi(2S)\to J/\psi+\pi\pi)$. Unlike the ISR process, $p_{J/\psi}$ ranges from $0$ to
$4.7$ GeV, and the momentum spectra of $J/\psi$ for the the t- and two-photon channel
parts are shown in Fig.[5].

\begin{figure}\label{2S_ee_angular}
\begin{center}
\subfigure[]
{\includegraphics[width=0.48\textwidth]{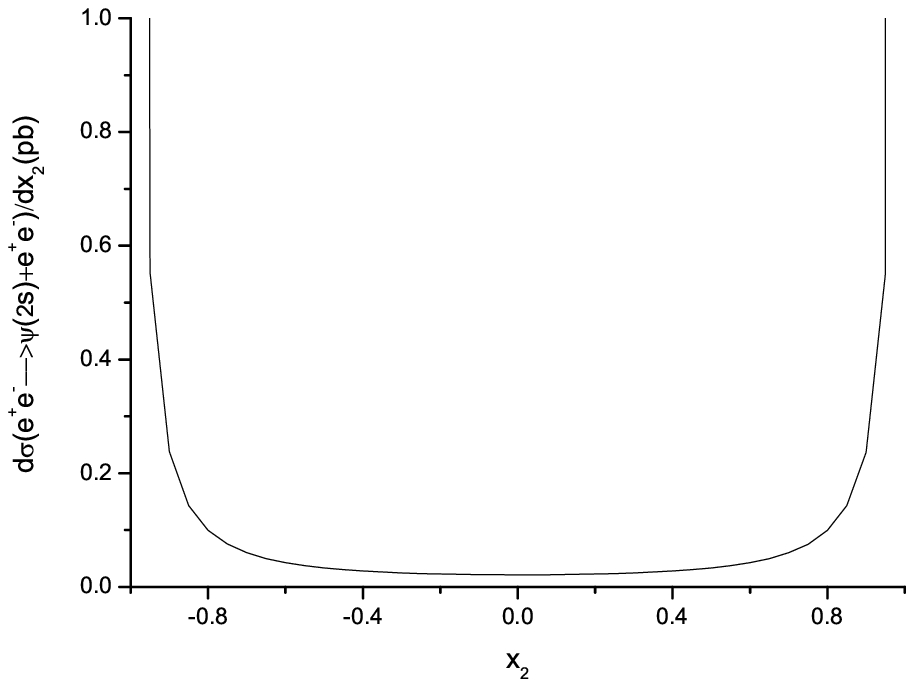}}
\subfigure[]
{\includegraphics[width=0.48\textwidth]{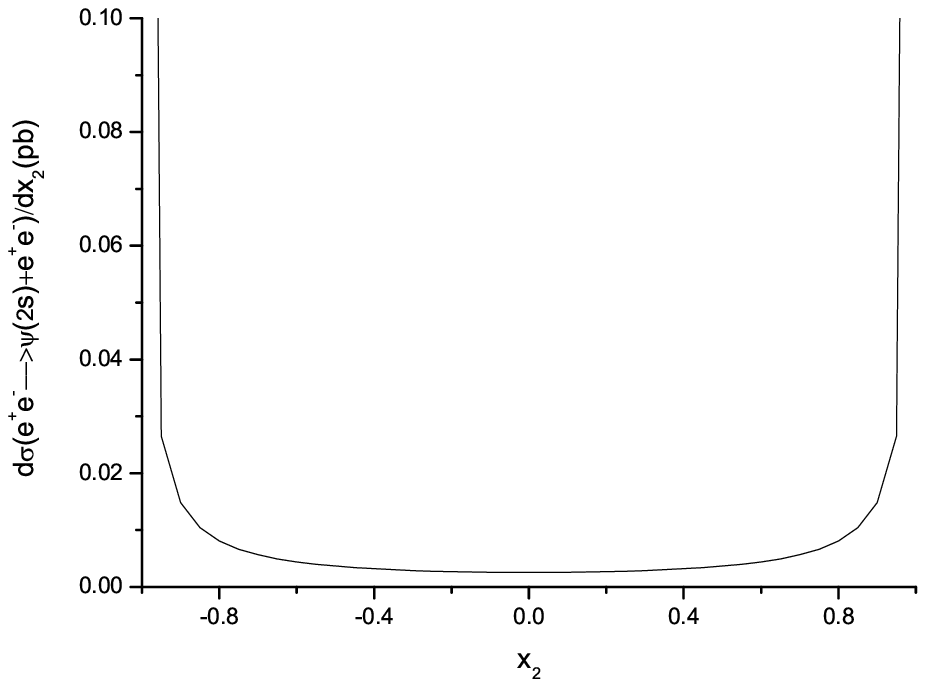}}
\caption{ The angular distribution of $\psi(2S)$ produced through t-channel (a), and
two-photon channel (b) in the $e^{+}e^{-}\to \psi(2S)+e^{+}e^{-}$ process, where
$x_2=\cos(\theta_{\psi(2S)})$, and $\theta_{\psi(2S)}$ is the angular between
$\psi(2S)$ and the $e^{+}e^{-}$ beam.}
\end{center}
\end{figure}

We also calculate the interference between the t-channel part and the two-photon part
and find it is very small. The cross section of the interference part for
$(J/\psi+\pi\pi)_{\psi(2S)}+e^{+}e^{-})$ is about $-20\mathrm{fb}$ in the whole phase
space region. After including the interference part, the total cross section in different
cut conditions of $\theta(J/\psi)$ are
\begin{subequations}\label{result6}
\begin{equation}
\sigma(e^{+}e^{-}\to (J/\psi+\pi\pi)_{\psi(2S)}+e^{+}e^{-})
\Big{|}_{\frac{\pi}{18}<\theta_{J/\psi}<\frac{17\pi}{18}}=0.12\mathrm{pb};
\end{equation}
\begin{equation}
\sigma(e^{+}e^{-}\to (J/\psi+\pi\pi)_{\psi(2S)}+e^{+}e^{-})
\Big{|}_{\frac{\pi}{9}<\theta_{J/\psi}<\frac{8\pi}{9}}=0.070\mathrm{pb};
\end{equation}
\begin{equation}
\sigma(e^{+}e^{-}\to (J/\psi+\pi\pi)_{\psi(2S)}+e^{+}e^{-})
\Big{|}_{\frac{\pi}{6}<\theta_{J/\psi}<\frac{5\pi}{6}}=0.047\mathrm{pb}.
\end{equation}
\end{subequations}
The angular distribution of $\psi(2S)$ and momentum distribution of $J/\psi$ for
the whole process can be approximately obtained by adding the t- and two-photon channel
contribution together respectively, because the interference effect is very small.

\begin{figure}\label{2S_ee_momentum}
\begin{center}
\subfigure[]
{\includegraphics[width=0.48\textwidth]{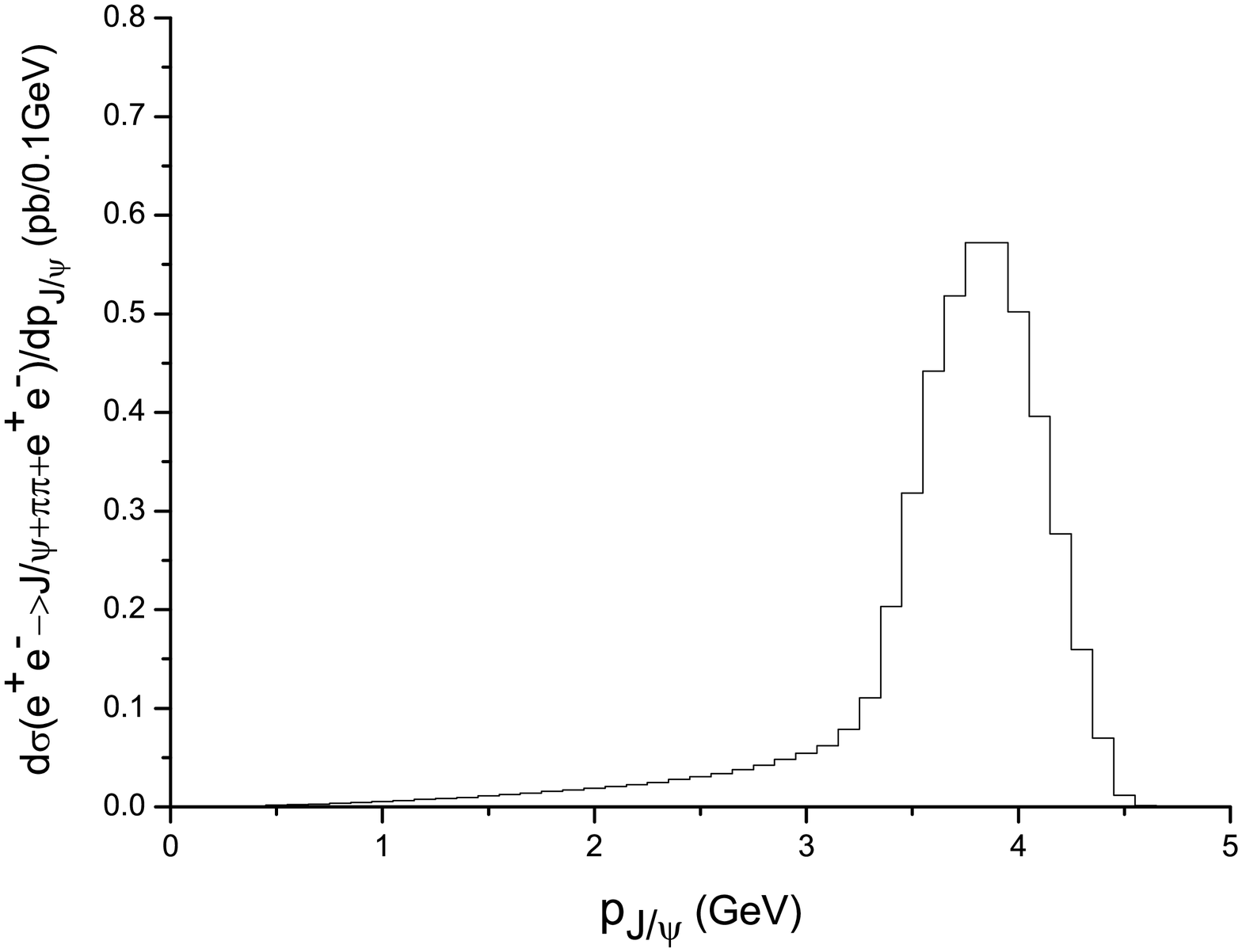}}
\subfigure[]
{\includegraphics[width=0.48\textwidth]{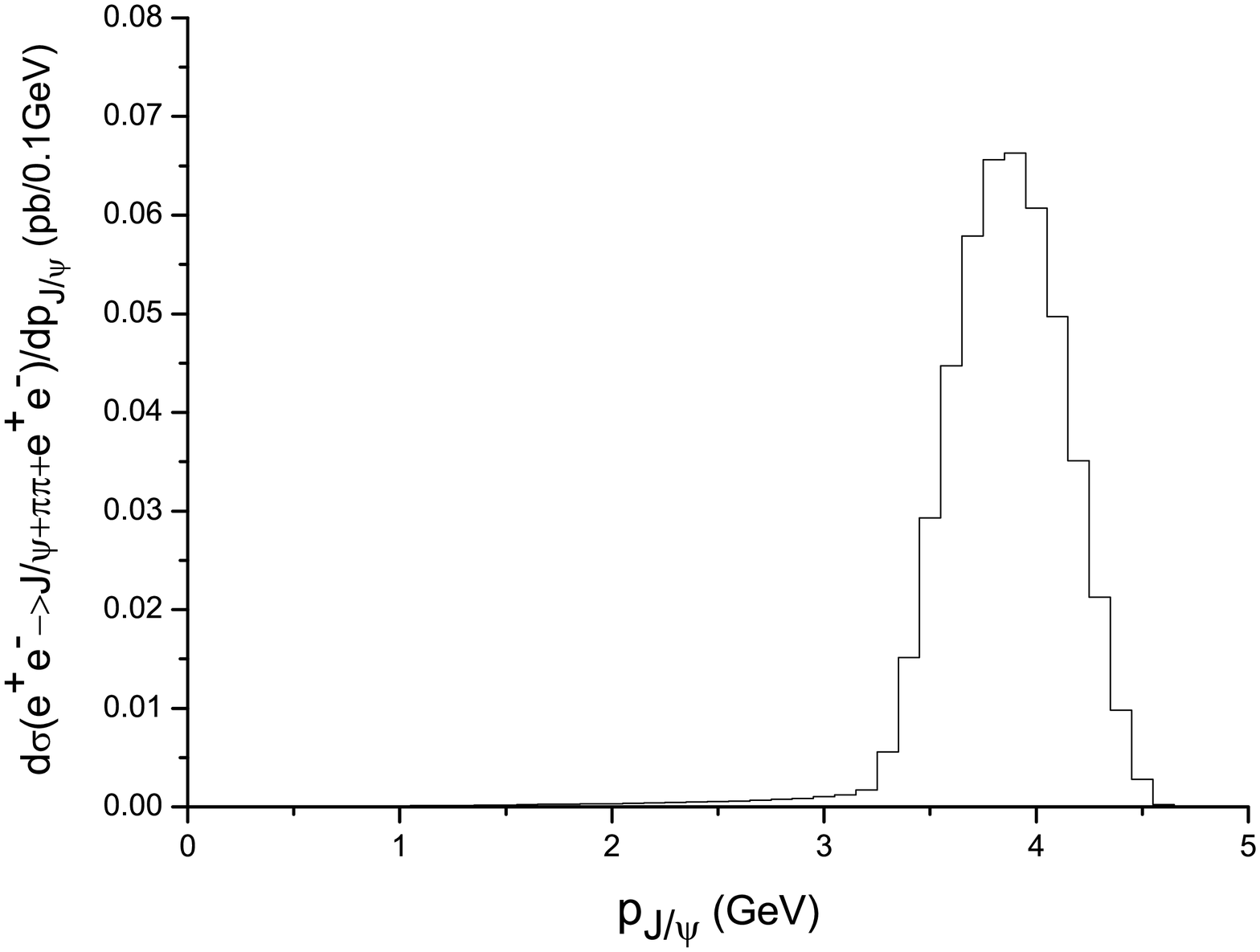}}
\caption{ The momentum distribution of $J/\psi$ produced through the t-channel (a)
and two-photon channel(b) in the $e^{+}e^{-}\to \psi(2S)+e^{+}e^{-}$ process.}
\end{center}
\end{figure}

\subsection{The Feed-Down Background From $e^{+}e^{-} \to \psi(2S)+f\bar{f}\;(f\neq e)$  }

  The process $e^{+}e^{-} \to \psi(2S)+f\bar{f}\;(f\neq e)$ has been fully studied in
Ref.\cite{Zhang:2010uia}. We also compute it independently and obtain consistent results:
\begin{equation}
\sum_{f=\mu,\tau,u,d,s}\sigma(e^{+}e^{-} \to (J/\psi+\pi\pi)_{\psi(2S)}+f\bar{f})=
0.026\mathrm{pb}.
\end{equation}
If we make the same cut on the $\theta_{J/\psi}$, the cross section becomes:
\begin{subequations}\label{result7}
\begin{equation}
\sum_{f=\mu,\tau,u,d,s}\sigma(e^{+}e^{-} \to (J/\psi+\pi\pi)_{\psi(2S)}+f\bar{f})
\Big{|}_{\frac{\pi}{18}<\theta_{J/\psi}<\frac{17\pi}{18}}=0.020\mathrm{pb};
\end{equation}
\begin{equation}
\sum_{f=\mu,\tau,u,d,s}\sigma(e^{+}e^{-} \to (J/\psi+\pi\pi)_{\psi(2S)}+f\bar{f})
\Big{|}_{\frac{\pi}{9}<\theta_{J/\psi}<\frac{8\pi}{9}}=0.015\mathrm{pb};
\end{equation}
\begin{equation}
\sum_{f=\mu,\tau,u,d,s}\sigma(e^{+}e^{-} \to (J/\psi+\pi\pi)_{\psi(2S)}+f\bar{f})
\Big{|}_{\frac{\pi}{6}<\theta_{J/\psi}<\frac{5\pi}{6}}=0.011\mathrm{pb}.
\end{equation}
\end{subequations}
The cross sections in different cut regions are within only about $0.020\mathrm{pb}$, which
are about $4\simeq6$ times less than those in the $e^{+}e^{-} \to \psi(2S)+e^{+}e^{-}$
process, so small that we will not present further analysis here, and recommend
Ref.\cite{Zhang:2010uia} for more detailed results.

\subsection{ The Background From $e^{+}e^{-} \to J/\psi+q\bar{q}$ }

The Feynman diagrams for the $e^{+}e^{-} \to J/\psi+q\bar{q}$ process are similar to those
for $e^{+}e^{-} \to \psi(2S)+q\bar{q}$ process. Since in the $\psi(2S)$ production process
the contribution of the s-channel diagrams can be ignored, for the same reason, we will not
consider it here too. The cross section of the $e^{+}e^{-} \to J/\psi+q\bar{q}$ has been
calculated in Ref.\cite{Zhang:2010uia}, which is also considerable. Using the method introduced
in Ref\cite{Zhang:2010uia}, we calculate the cross section with different cut conditions of
$\theta_{J/\psi}$:
\begin{subequations}\label{result8}
\begin{equation}
\sum_{q=u,d,s}\sigma(e^{+}e^{-}\to J/\psi+q\bar{q})
\Big{|}_{\frac{\pi}{18}<\theta_{J/\psi}<\frac{17\pi}{18}}=0.071\mathrm{pb};
\end{equation}
\begin{equation}
\sum_{q=u,d,s}\sigma(e^{+}e^{-}\to J/\psi+q\bar{q})
\Big{|}_{\frac{\pi}{9}<\theta_{J/\psi}<\frac{8\pi}{9}}=0.052\mathrm{pb};
\end{equation}
\begin{equation}
\sum_{q=u,d,s}\sigma(e^{+}e^{-}\to J/\psi+q\bar{q})
\Big{|}_{\frac{\pi}{6}<\theta_{J/\psi}<\frac{5\pi}{6}}=0.039\mathrm{pb}.
\end{equation}
\end{subequations}
The $J/\psi$ angular and momentum distribution are shown in Fig.[6].
Note the difference between our results and those in Ref.\cite{Zhang:2010uia} is due to
the different choice of the parameters and the amount of data samples used in the $R$-value
curve \cite{Nakamura:2010zzi}.

\begin{figure}
\begin{center}\label{jsi_qq}
\subfigure[]
{\includegraphics[width=0.48\textwidth]{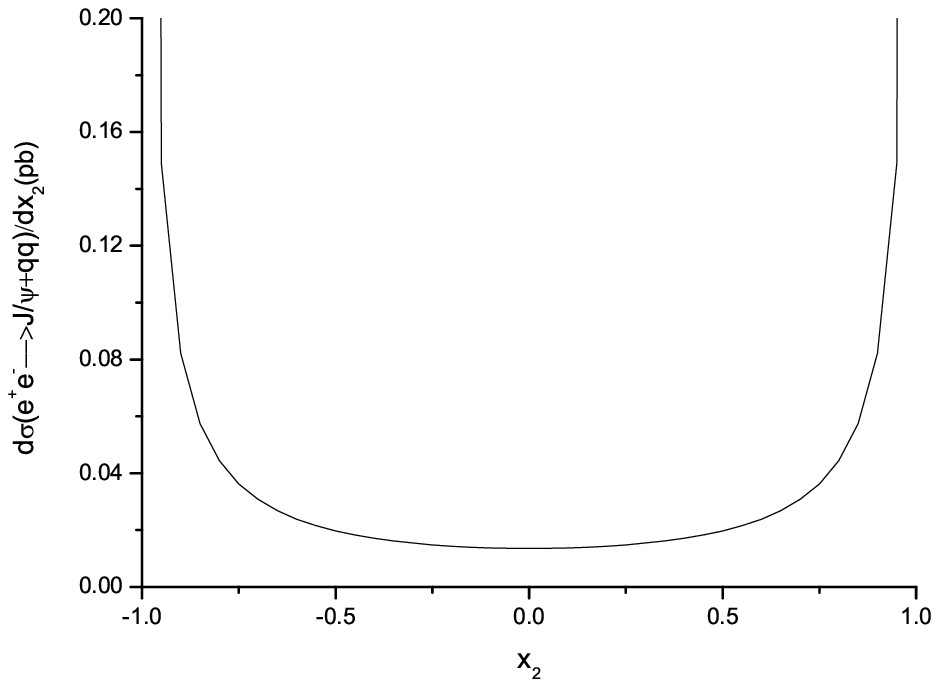}}
\subfigure[]
{\includegraphics[width=0.48\textwidth]{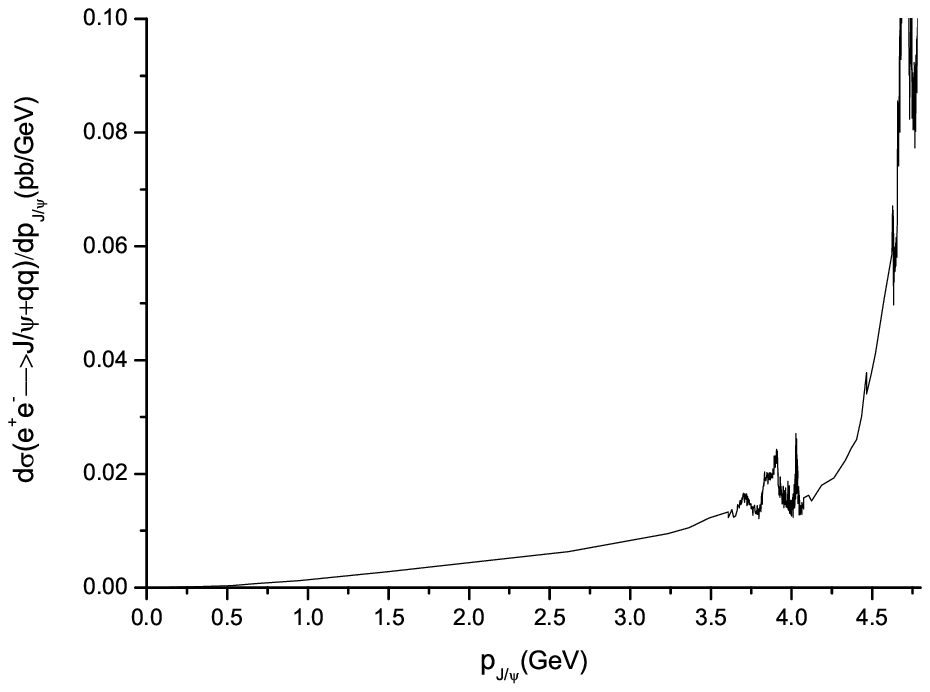}}
\caption{ The angular (a) and momentum (b) distributions of $J/\psi$ in the process of
$e^{+}e^{-}\to J/\psi+q\bar{q}$, where $x^{\prime}_2=\cos(\theta_{J/\psi})$,
and $\theta_{J/\psi}$ is the angular between $\theta_{J/\psi}$ and the $e^{+}e^{-}$ beam.}
\end{center}
\end{figure}

\section{Discussions and Conclusions}

  Summing up the feed-down contribution from the ISR and $f\bar{f}$ processes and the
contribution of direct $J/\psi+q\bar{q}$ production, the total QED background cross section
are about
\begin{equation}
\sigma_{QED}(e^{+}e^{-}\to J/\psi+ LH)=7.46\mathrm{pb},
\end{equation}
which is more than one order of magnitude larger than the cross section of
the conventional QCD production $e^{+}e^{-}\to J/\psi+\mathrm{LH}$
\cite{Cho:1996cg,Ma:2008gq,Gong:2009kp,Zhang:2009ym}.
Such huge background make it difficult to measure the QCD contribution in the whole phase
space region. However, the background in the off beam region will drop down deeply.
The cross section of the QED background in different cut regions are:
\begin{subequations}\label{result8}
\begin{equation}
\sigma_{QED}(e^{+}e^{-}\to J/\psi+ LH)
\Big{|}_{\frac{\pi}{18}<\theta_{J/\psi}<\frac{17\pi}{18}}=1.73\mathrm{pb};
\end{equation}
\begin{equation}
\sigma_{QED}(e^{+}e^{-}\to J/\psi+ LH)
\Big{|}_{\frac{\pi}{9}<\theta_{J/\psi}<\frac{8\pi}{9}}=1.14\mathrm{pb};
\end{equation}
\begin{equation}
\sigma_{QED}(e^{+}e^{-}\to J/\psi+ LH)
\Big{|}_{\frac{\pi}{6}<\theta_{J/\psi}<\frac{5\pi}{6}}=0.81\mathrm{pb}.
\end{equation}
\end{subequations}

In NRQCD, the conventional $J/\psi+\mathrm{LH}$ production includes the both the CS and
the CO contribution. For the CS process $e^{+}e^{-}\to J/\psi+gg$, both the NLO QCD
corrections\cite{Ma:2008gq} and relativistic corrections\cite{He:2009uf,Jia:2009np} have
been calculated. The cross section is found to be $0.4\sim0.7\;\mathrm{pb}$ at NLO in
$\alpha_s$ and $v_c^2$ \cite{Ma:2008gq,Gong:2009kp,He:2009uf}. The NLO QCD corrections
to the CO contribution have also been obtained \cite{Zhang:2009ym}. If we choose
$\langle0|\mathcal{O}(^1S_0^{8})|0\rangle^{J/\psi}=(3.04\pm0.35)\times10^{-2}\rm{GeV}^{3},\;
\langle0|\mathcal{O}(^3P_J^{8})|0\rangle^{J/\psi}=(-9.08\pm1.61)\times10^{-3}\rm{GeV}^{5}$,
which are obtained by a global fitting of $J/\psi$ production data\cite{Butenschoen:2011yh},
the cross section of the CO contribution at $\alpha_s$ NLO will be about $0.3\mathrm{pb}$
at $\mu=2m_c,\alpha_s({\mu})=0.245$. Then the total NRQCD prediction for the conventional
$J/\psi+\mathrm{LH}$ production will be about $0.7\sim1.0\mathrm{pb}$. Unlike the QED
background, the cut on $\theta_{J/\psi}$, for example $\pi/9<\theta_{J/\psi}<8\pi/9$,
will only have a minor influence on the conventional QCD cross section $\sigma^{QCD}$, because
both the CS and CO contribution do not depend strongly on $\theta_{J/\psi}$ \cite{Braaten:1995ez,Gong:2009kp}.
Therefore, we conclude that in a suitable cut condition of $\theta_{J/\psi}$, the cross section
of the conventional QCD process can be in the same order as the background cross section.
Furthermore, the results in \cite{Zhang:2009ym} shown that CO contribution mainly assemble in
the kinematic end point region, while the CS contribution is distributed in the whole region
of $0<p_{J/\psi}<4.85\mathrm{GeV}$, so to study the CO contribution, it can be further required
$p_{J/\psi}>3\mathrm{GeV}$ in the measurement. Such a requirement will reduced the CS contribution
by about $50\%$, but has little affect on the CO and the QED background contribution.
In our calculation, we determine the effective vertices of $\psi(2S)\gamma^{\ast}$,
$J/\psi\gamma^{\ast}$ and $\psi(2S)\to J/\psi+\pi\pi$ by fitting the experimental data and
using the $R$-value to represent the effective vertex of $\gamma^{\ast} q \bar{q}$ in the calculation
of $e^{+}e^{-} \to J/\psi(\psi(2S))+q\bar{q}$ cross section, this indicates that all the possible
important higher QCD correction effects to the background are included automatically, which
makes the uncertainties of our result very small. Based on the above analysis, we think further
measurement of the $J/\psi+\mathrm{LH}$ production with a suitable cut condition of $\theta_{J/\psi}$
and $p_{J/\psi}$ will be helpful to understand the role of the CO contribution to the $J/\psi$
production mechanism in $e^{+}e^{-}$ annihilation.

Recently the complete NLO QCD correction to the polarization of $J/\psi$ hadroproduction were
obtained by two groups \cite{Butenschoen:2012px,Chao:2012iv}. Due to their different ways of
fitting the CO matrix elements, they got completely different conclusions. After taking into
account the feed-down contribution of $\chi_{cJ}$ and $\psi(2S)$ states \cite{Gong:2012ug}, the
authors found that there is no solution to fit the $p_t$ distribution of the cross section and
$J/\psi$ polarization measured by CDF collaboration simultaneously. Understanding the $J/\psi$ production at
B-factories can also help to resolve the polarization problem of $J/\psi$ hadroproduction.

In summary, we study the dominant background sources of $J/\psi+\mathrm{LH}$ production in
$e^{+}e^{-}$ annihilation, which include the ISR process $e^{+}e^{-}\to \psi(2S)+\gamma$ and
higher QED process $e^{+}e^{-}\to \psi(2S)+f\bar{f}$, where $f$ can be lepton or light quark,
as well as the direct $e^{+}e^{-}\to J/\psi+q\bar{q}$ process with $q=u,d,s$ quark.
We find that the cross section of the background process is very large in the whole phase
space region. If we make a cut on the angle between $J/\psi$ and $e^{+}e^{-}$ beam,
the cross section of the QED processes will reduced largely and become comparable to the
cross section of the conventional QCD process. This indicates it is possible to measure
the cross section of $J/\psi+\mathrm{LH}$ from conventional QCD production at B-factories.

\section*{Acknowledgments}
We would like to thank ChangZheng Yuan and  Chao-Hsi Chang for helpful discussion.
Z.G.He thanks the Theoretical Physics Center for Science Facilities(CAS) and
Universitat de Barcelona for hospitality while part of this work was carried out. This work
is supported by the National Natural Science Foundation of China (No. 10979056 and No.10935012),
and by the Chinese Academy of Science under Project No. INFO-115-B01. The work of Z.G.He
was supported in part by the CSD2007-00042 Consolider-Ingenio 2010 program under the
CPAN08-PD14 contract, by the FPA2007-66665-C02-01/ and FPA2010-16963 projects (Spain),
and is supported by the German Federal Ministry for Education and Research BMBF through 
Grant No. 05H12GUE.








\begin{thebibliography}{99}

\bibitem{Bodwin:1994jh}
  G.~T.~Bodwin, E.~Braaten and G.~P.~Lepage,
  Phys.\ Rev.\  D {\bf 51}, 1125 (1995)
  [Erratum-ibid.\  D {\bf 55}, 5853 (1997)].


\bibitem{Brambilla:2004wf}
  N.~Brambilla {\it et al.}  [Quarkonium Working Group],
  arXiv:hep-ph/0412158;
  N.~Brambilla {\it et al.},
  Eur.\ Phys.\ J.\  C {\bf 71}, 1534 (2011).


\bibitem{Aubert:2001pd}
  B.~Aubert {\it et al.}  [BABAR Collaboration],
  Phys.\ Rev.\ Lett.\  {\bf 87}, 162002 (2001).


\bibitem{Abe:2001za}
  K.~Abe {\it et al.}  [BELLE Collaboration],
  Phys.\ Rev.\ Lett.\  {\bf 88}, 052001 (2002).


\bibitem{Abe:2002rb}
  K.~Abe {\it et al.}  [Belle Collaboration],
  Phys.\ Rev.\ Lett.\  {\bf 89}, 142001 (2002);
  K.~Abe {\it et al.}  [Belle Collaboration],
  Phys.\ Rev.\  D {\bf 70}, 071102 (2004);


\bibitem{:2009nj}
  P.~Pakhlov {\it et al.}  [Belle Collaboration],
  Phys.\ Rev.\  D {\bf 79}, 071101 (2009).


\bibitem{Aubert:2005tj}
  B.~Aubert {\it et al.}  [BABAR Collaboration],
  Phys.\ Rev.\  D {\bf 72}, 031101 (2005).


\bibitem{Cho:1996cg}
  P.~L.~Cho and A.~K.~Leibovich,
  Phys.\ Rev.\  D {\bf 54}, 6690 (1996);
  S.~Baek, P.~Ko, J.~Lee and H.~S.~Song,
  J.\ Korean Phys.\ Soc.\  {\bf 33}, 97 (1998);


\bibitem{Liu:2003jj}
  K.~Y.~Liu, Z.~G.~He and K.~T.~Chao,
  Phys.\ Rev.\  D {\bf 69}, 094027 (2004).


\bibitem{Braaten:2002fi}
  E.~Braaten and J.~Lee,
  Phys.\ Rev.\  D {\bf 67}, 054007 (2003)
  [Erratum-ibid.\  D {\bf 72}, 099901 (2005)];
  K.~Y.~Liu, Z.~G.~He and K.~T.~Chao,
  Phys.\ Lett.\  B {\bf 557}, 45 (2003).


\bibitem{Zhang:2005cha}
  Y.~J.~Zhang, Y.~j.~Gao and K.~T.~Chao,
  Phys.\ Rev.\ Lett.\  {\bf 96}, 092001 (2006);
  Y.~J.~Zhang and K.~T.~Chao,
  Phys.\ Rev.\ Lett.\  {\bf 98}, 092003 (2007);
  Y.~J.~Zhang, Y.~Q.~Ma and K.~T.~Chao,
  Phys.\ Rev.\  D {\bf 78}, 054006 (2008).


\bibitem{Gong:2007db}
  B.~Gong and J.~X.~Wang,
  Phys.\ Rev.\  D {\bf 77}, 054028 (2008);
  B.~Gong and J.~X.~Wang,
  Phys.\ Rev.\ Lett.\  {\bf 100}, 181803 (2008);
  B.~Gong and J.~X.~Wang,
  Phys.\ Rev.\  D {\bf 80}, 054015 (2009).



\bibitem{Bodwin:2006ke}
  G.~T.~Bodwin, D.~Kang, T.~Kim, J.~Lee and C.~Yu,
  AIP Conf.\ Proc.\  {\bf 892}, 315 (2007);
  Z.~G.~He, Y.~Fan and K.~T.~Chao,
  Phys.\ Rev.\  D {\bf 75}, 074011 (2007).


\bibitem{Braaten:1995ez}
  E.~Braaten and Y.~Q.~Chen,
  Phys.\ Rev.\ Lett.\  {\bf 76}, 730 (1996);


\bibitem{Yuan:1996ep}
  F.~Yuan, C.~F.~Qiao and K.~T.~Chao,
  Phys.\ Rev.\  D {\bf 56}, 321 (1997);


\bibitem{Fleming:2003gt}
  S.~Fleming, A.~K.~Leibovich and T.~Mehen,
  Phys.\ Rev.\  D {\bf 68}, 094011 (2003);


\bibitem{Wang:2003fw}
  J.~X.~Wang,
  arXiv:hep-ph/0311292;


\bibitem{Driesen:1993us}
  V.~M.~Driesen, J.~H.~Kuhn and E.~Mirkes,
  Phys.\ Rev.\  D {\bf 49}, 3197 (1994).


\bibitem{Ma:2008gq}
  Y.~Q.~Ma, Y.~J.~Zhang and K.~T.~Chao,
  Phys.\ Rev.\ Lett.\  {\bf 102}, 162002 (2009);

\bibitem{Gong:2009kp}
  B.~Gong and J.~X.~Wang,
  Phys.\ Rev.\ Lett.\  {\bf 102}, 162003 (2009).


\bibitem{Zhang:2009ym}
  Y.~J.~Zhang, Y.~Q.~Ma, K.~Wang and K.~T.~Chao,
  Phys.\ Rev.\  D {\bf 81}, 034015 (2010).


\bibitem{He:2009uf}
  Z.~G.~He, Y.~Fan and K.~T.~Chao,
  Phys.\ Rev.\  D {\bf 81}, 054036 (2010).


\bibitem{Jia:2009np}
  Y.~Jia,
  Phys.\ Rev.\  D {\bf 82}, 034017 (2010).


\bibitem{Li:2010xu}
  R.~Li and J.~X.~Wang,
  Phys.\ Rev.\  D {\bf 82}, 054006 (2010).

\bibitem{He:2009by}
  Z.~G.~He and J.~X.~Wang,
  Phys.\ Rev.\  D {\bf 81}, 054030 (2010);
  Z.~G.~He and J.~X.~Wang,
  Phys.\ Rev.\  D {\bf 82}, 094033 (2010).


\bibitem{Artoisenet:2009xh}
  P.~Artoisenet, J.~M.~Campbell, F.~Maltoni and F.~Tramontano,
  Phys.\ Rev.\ Lett.\  {\bf 102}, 142001 (2009).


\bibitem{Chang:2009uj}
  C.~H.~Chang, R.~Li and J.~X.~Wang,
  Phys.\ Rev.\  D {\bf 80}, 034020 (2009).


\bibitem{Butenschoen:2009zy}
  M.~Butenschoen and B.~A.~Kniehl,
  Phys.\ Rev.\ Lett.\  {\bf 104}, 072001 (2010).


\bibitem{Campbell:2007ws}
  J.~M.~Campbell, F.~Maltoni and F.~Tramontano,
  Phys.\ Rev.\ Lett.\  {\bf 98}, 252002 (2007).


\bibitem{Gong:2008sn}
  B.~Gong and J.~X.~Wang,
  Phys.\ Rev.\ Lett.\  {\bf 100}, 232001 (2008).


\bibitem{Gong:2008hk}
  B.~Gong and J.~X.~Wang,
  Phys.\ Rev.\  D {\bf 78}, 074011 (2008).


\bibitem{Braaten:1994vv}
  E.~Braaten and S.~Fleming,
  Phys.\ Rev.\ Lett.\  {\bf 74}, 3327 (1995).


\bibitem{Gong:2008ft}
  B.~Gong, X.~Q.~Li and J.~X.~Wang,
  Phys.\ Lett.\  B {\bf 673}, 197 (2009).


\bibitem{Ma:2010yw}
  Y.~Q.~Ma, K.~Wang and K.~T.~Chao,
  Phys.\ Rev.\ Lett.\  {\bf 106}, 042002 (2011).


\bibitem{Butenschoen:2010rq}
  M.~Butenschoen and B.~A.~Kniehl,
  Phys.\ Rev.\ Lett.\  {\bf 106}, 022003 (2011).


\bibitem{Benayoun:1999hm}
  M.~Benayoun, S.~I.~Eidelman, V.~N.~Ivanchenko and Z.~K.~Silagadze,
  Mod.\ Phys.\ Lett.\  A {\bf 14}, 2605 (1999).

\bibitem{Chang:1997dw}
  C.~H.~Chang, C.~F.~Qiao and J.~X.~Wang,
  Phys.\ Rev.\  D {\bf 56}, 1363 (1997);
\bibitem{Chang:1998pz}
  C.~H.~Chang, C.~F.~Qiao and J.~X.~Wang,
  Phys.\ Rev.\  D {\bf 57}, 4035 (1998).


\bibitem{Mannel:1995jt}
  T.~Mannel and R.~Urech,
  Z.\ Phys.\  C {\bf 73}, 541 (1997).


\bibitem{Bai:1999mj}
  J.~Z.~Bai {\it et al.}  [BES Collaboration],
  Phys.\ Rev.\  D {\bf 62}, 032002 (2000).


\bibitem{Nakamura:2010zzi}
  K.~Nakamura {\it et al.}  [Particle Data Group],
  J.\ Phys.\ G {\bf 37}, 075021 (2010).

\bibitem{Wang:2004du}
  J.~X.~Wang,
  Nucl.\ Instrum.\ Meth.\  A {\bf 534}, 241 (2004).


\bibitem{Zhang:2010uia}
  Y.~J.~Zhang, B.~Q.~Li and K.~Y.~Liu,
  arXiv:1003.5566 [hep-ph].


\bibitem{Butenschoen:2011yh}
  M.~Butenschoen and B.~A.~Kniehl,
  arXiv:1105.0820 [hep-ph].

\bibitem{Butenschoen:2012px}
  M.~Butenschoen and B.~A.~Kniehl,
Phys.\ Rev.\ Lett.\  {\bf 108}, 172002 (2012).

\bibitem{Chao:2012iv}
  K.~-T.~Chao, Y.~-Q.~Ma, H.~-S.~Shao, K.~Wang and Y.~-J.~Zhang,
Phys.\ Rev.\ Lett.\  {\bf 108}, 242004 (2012).


\bibitem{Gong:2012ug}
  B.~Gong, L.~-P.~Wan, J.~-X.~Wang and H.~-F.~Zhang,
arXiv:1205.6682 [hep-ph].


\end{thebibliography}
\end{document}